\documentclass[reprint,aip,jap,superscriptaddress,amsmath,amssymb,preprintnumbers,showpacs]{revtex4-1}

\usepackage{graphicx}
\usepackage{dcolumn}
\usepackage{bm}
\usepackage{times}
\usepackage{subfigure}
\usepackage{hyperref}
\usepackage{natbib}
\usepackage{amsmath}
\usepackage{booktabs}
\usepackage{comment}
\newcommand{\eqn}[2]{\begin{equation}\label{#2}#1\end{equation}}
\newcommand{\mt}[1]{\mathrm{#1}}

\begin{document}

\title{Validation of the Wiedemann-Franz Law in solid and molten tungsten above 2000 K through thermal conductivity measurements via steady state temperature differential radiometry}
\author{Milena Milich}
\thanks{These two authors contributed equally}
\affiliation{Department of Mechanical and Aerospace Engineering, University of Virginia, Charlottesville, VA, USA 22904}
\author{Hunter B. Schonfeld}
\thanks{These two authors contributed equally}
\affiliation{Department of Mechanical and Aerospace Engineering, University of Virginia, Charlottesville, VA, USA 22904}
\author{Konstantinos Boboridis}
\affiliation{European Commission, Joint Research Centre, Karlsruhe, Germany D-76125}
\author{Davide Robba}
\affiliation{European Commission, Joint Research Centre, Karlsruhe, Germany D-76125}
\author{Luka Vlahovic}
\affiliation{European Commission, Joint Research Centre, Karlsruhe, Germany D-76125}
\author{Rudy Konings}
\affiliation{European Commission, Joint Research Centre, Karlsruhe, Germany D-76125}
\author{Jeffrey L. Braun}
\affiliation{Laser Thermal, Inc., Charlottesville, VA, USA 22902}
\author{John T. Gaskins}
\affiliation{Laser Thermal, Inc., Charlottesville, VA, USA 22902}
\author{Niraj Bhatt}
\affiliation{Mechanical, Industrial and Systems Engineering, University of Rhode Island, Kingston, RI, USA 02881}
\author{Ashutosh Giri}
\affiliation{Mechanical, Industrial and Systems Engineering, University of Rhode Island, Kingston, RI, USA 02881}
\author{Patrick E. Hopkins}
\email{phopkins@virginia.edu}
\affiliation{Department of Mechanical and Aerospace Engineering, University of Virginia, Charlottesville, VA, USA 22904}
\affiliation{Department of Materials Science and Engineering, University of Virginia, Charlottesville, VA, USA 22904}
\affiliation{Department of Physics, University of Virginia, Charlottesville, VA, USA 22904}

\begin{abstract}
We measure the thermal conductivity of solid and molten tungsten using Steady State Temperature Differential Radiometry.  We demonstrate that the thermal conductivity can be well described by application of Wiedemann-Franz Law to electrical resistivity data, thus suggesting the validity of Wiedemann-Franz Law to capture the electronic thermal conductivity of metals in their molten phase.  We further support this conclusion using \textit{ab initio} molecular dynamics simulations with a machine-learned potential.  Our results show that at these high temperatures, the vibrational contribution to thermal conductivity is negligible compared to the electronic component.
\end{abstract}

\maketitle

The critical heat transfer mechanisms of materials subjected to extreme heat fluxes and temperatures are rooted in an interplay between conduction and radiation.  For metals in particular, the combination of their high thermal conductivities and low emissivities make the conduction of heat away from heated surfaces a relatively substantial energy transfer process at thousands of Kelvin, especially for high melting point refractory metals.\cite{ho1972aa,touloukian1970ad,allen1960aa}  This makes refractory metals ideal materials for use in nuclear fusion and additive manufacturing processes, for example, due to their ability to withstand the large temperature changes in reactors or extreme heat fluxes imparted by laser heating, respectively.  A key thermal transport property of metals for these high heat flux applications is thus the thermal conductivity, and due to the large resultant temperature excursions concerns both solid and molten phases.  However, measurements of of thermal conductivity during these extreme conditions are particularly challenging.  This has led to a void in the fundamental understanding of the conductive processes of metals at high temperatures and in their molten phases resulting in century-old theories such as the Wiedemann-Franz Law for electronic thermal conductivity\cite{wiedemann1853aa} unvalidated for metals in this regime, and in particular in their molten state.

In this work, we report on a new measurement technique that we develop to measure the thermal conductivity of materials as a function of temperature at elevated temperatures including into their molten states.  We measure the thermal conductivity of W from $\sim 2000-4000$ K using an approach we call ``Steady State Temperature Differential Radiometry'' (SSTDR), in which we heat the sample with a continuous wave laser source to induce steady state conditions, and then measure the resulting temperature changes as a function of laser power perturbations via radiative pyrometry, which we then relate to the thermal conductivity via the Fourier Law.  By the nature of these SSTDR measurements being conducted in near steady state conditions, the technique is a measurement of thermal conductivity, as opposed to an effusivity or diffusivity measurements typical using transient or frequency domain techniques, thus avoiding the requirement of knowledge of the material's heat capacity to deduce the thermal conductivity.  The nature of this technique being single sided (heat and measure temperature on the same side of the sample) and the maximally heated volume being constrained to the laser spot size allow us to induce a sustained molten pool in the middle of an otherwise solid sample, thus enabling a versatile technique for repeated measurement of thermal conductivity and total emissivity of materials in their molten states.  We use this to measure the thermal conductivity, $k$, of molten W, an exemplary metal with the highest melting point of any elemental solid with extensively studied high temperature radiative properties allowing for a judiciously selected material for validation of SSTDR measurements.\cite{ho1972aa,touloukian1970ad,allen1960aa,sievers1978aa,grimvall1987aa,wang2011ad,chen2019aa,pottlacher2010aa,tolias2017aa,matsumoto1999aa,bedford1996aa,cezairliyan1992aa,cezairliyan1971aa,cezairliyan1996aa,miiller1993aa,larrabee1959aa,brodu2015aa} We validate this technique via measurements of $k$ of solid W from $2000$ K up to its melting point with agreement with prior literature,\cite{ho1972aa,tolias2017aa,touloukian1970ad} further supporting the utility of this technique.  Extending the measurement to molten W allows us to demonstrate that the thermal conductivity can be well described by application of Wiedemann-Franz Law to electrical resistivity data, thus also suggesting the validity of Wiedemann-Franz Law to capture the electronic thermal conductivity of metals in their molten phase.  We further support this conclusion by calculating the phonon/vibrational contribution to thermal conductivity of W up through its melting point using \textit{ab initio} molecular dynamics (MD) simulations, were we develop a machine-learned (ML) potential for W that is uniformly accurate from room temperature through the melting point.  Our results show that at these high temperatures, the vibrational contribution to thermal conductivity is negligible compared to the electronic component, thus demonstrating that the Wiedemann-Franz Law can give an accurate prediction of the thermal conductivity of W at high temperatures, including its molten state.

\begin{figure}[t] 
\begin{center}
\includegraphics[scale=.17]{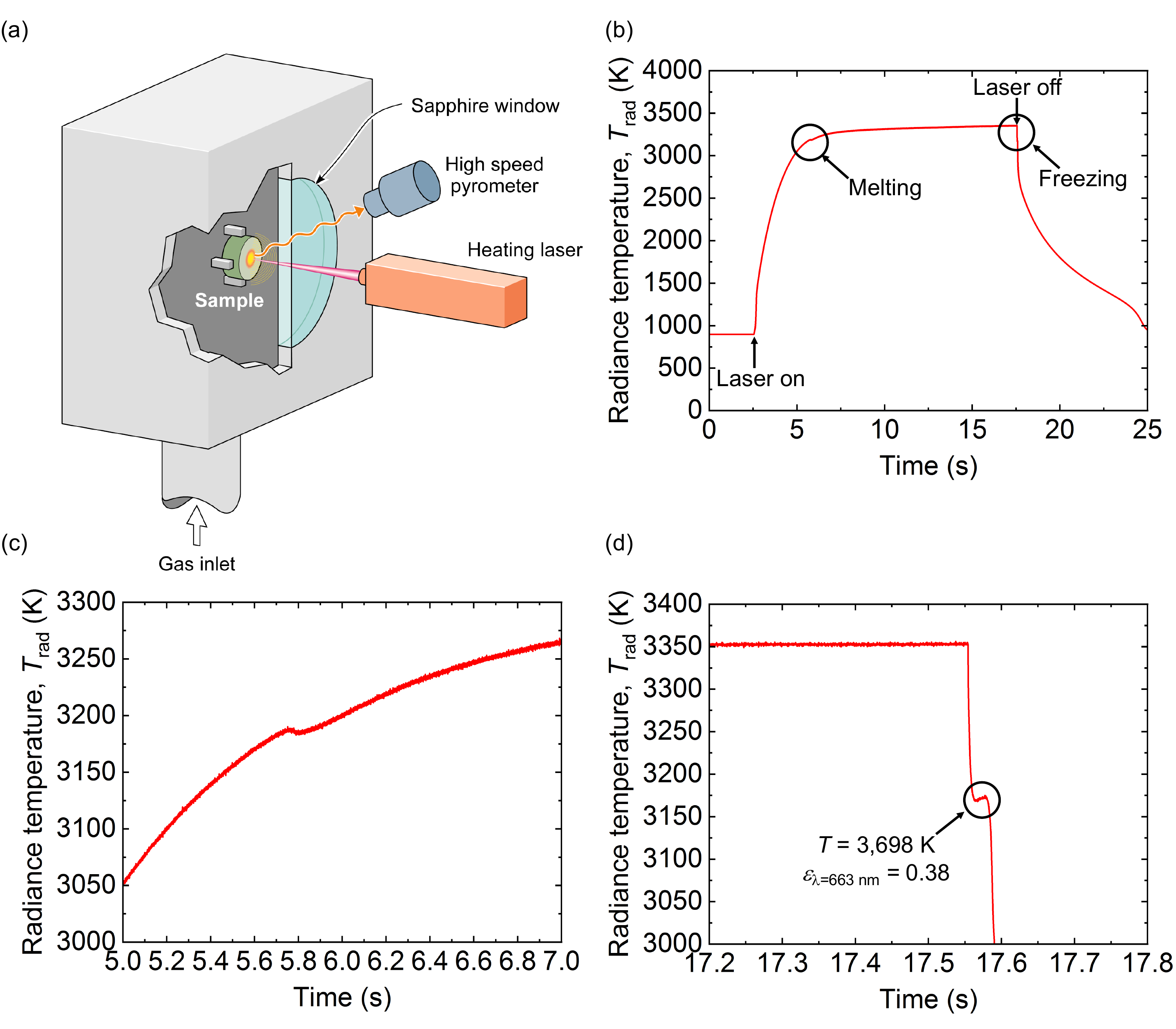}
\caption{(a) Schematic of experimental set up.  A key in this approach is the ability to locally heat a region of the sample to its molten state while pyrometrically sensing radiance temperature in the middle of the heated region.  This ``self-crucible'' approach then allows for thermal conductivity measurements by just sustaining melt only the middle of the sample during quasi-steady state conditions. (b)  Typical thermogram of the W sample heated to achieve quasi-steady state conditions with a molten center, with the specific onset times of melting and resolidification (i.e., ``freezing'') indicated in the plot.  The incident laser power generating these radiance temperature was 1.9 kW.  The voltage response of the our particular pyrometer was not linear with temperature until above $\sim$1600 K radiance temperature, indicative of the non-physical cooling response for times greater than 20 s.  The minimum temperature that the pyrometer would register was $\sim$900 K radiance temperature, thus leading to this reading at all times when the sample was below this temperature. (c) Data from (b) zoomed in around the melting temperature. (d) Data from (b) zoomed in around the freezing temperature.  The temperature at which this thermal arrest occurs is used to determine the melting temperature of the sample.}
\label{Figexperiment}
\end{center}
\end{figure}

The details of the equipment used in our experimental set up is described in a prior work,\cite{manara2017aa,cedillos-barraza2016aa} which describes the principles and additional specific considerations of our equipment for pyrometric sensing of temperature of laser heated solids to measure their melting point.  We use this configuration, the pertinent equipment from which are shown in Fig.~\ref{Figexperiment}a, and modify the laser heating profile to measure the thermal conductivity. Our W sample is a cylinder that is 1 cm in diameter and 2 mm thick. We use a high power continuous wave laser (wavelength 1064 nm, maximum power 2 kW) focused through the sapphire window on the chamber (transmissivity 0.86) down to a 3.5 mm diameter with a top-hat intensity profile to heat the sample and pressurize the chamber to 1.5 bar in Ar to minimize convective losses.\cite{pavlov2018aa}  The temperature of the middle of the heated area on the surface of the W is measured with a high speed pyrometer at 663 nm with a 0.8 mm field of view.  We control the temperature of the measured region of the laser heated W by changing the laser power, and continue to heat the sample until quasi-steady state conditions are reached, as determined from a relatively constant temperature in time. Prior to measurements, we heat the W sample through its melting point and cool to ambient several times to effectively clean and smooth the W surface in the measurement volume to ensure consistent surface conditions among measurements. We are able to sustain the melt pool in the center of the W sample for several minutes, effectively demonstrating that this measurement approach uses the sample as a ``self-crucible'' with the volume of the sample outside of the heat affected zone remaining solid. 

In the exemplar data show in Fig.~\ref{Figexperiment}b, we heat the W sample through its melting point, and the molten volume reaches a quasi steady state about 10-15 seconds after initiation of laser heating.  The time it takes for the sample to reach these steady conditions is related to the thermal diffusivity and laser spot size, which we discuss in detail in our prior works.\cite{braun2019aa,braun2018aa} A clear thermal arrest occurs in our temperature vs.~time profile when the sample undergoes a phase change to/from its liquid state, shown in Fig.~\ref{Figexperiment}c and d.  As detailed previously, we use thermal arrest associated with the resolidification (i.e., ``freezing'') of the W after the sample turns off.  Using Wien's Law, the true temperature can be calculated from the radiance temperature, $T_{\mt{rad}}$ from
\eqn{\frac{1}{T}=\frac{1}{T_{\mt{rad}}}+\frac{\lambda}{c_2}\ln \varepsilon_{\lambda}}{eqWien}
where $\lambda$ is the wavelength, $\varepsilon_{\lambda}$ is the spectral emissivity at $\lambda$, and $c_2= 14388$ $\mu$m-K, from which we measure the melting temperature of our W sample as $T_{\mt{melt}}=3698$ K using $\varepsilon_{\lambda=663\,\mt{nm}} = 0.38$ (Fig.~\ref{Figexperiment}d) from accepted values in the literature,\cite{touloukian1970ad} in excellent agreement with the literature ($T_{\mt{melt}}=3,707$ K).\cite{lide2008aa}

\begin{figure}[t] 
\begin{center}
\includegraphics[scale=.4]{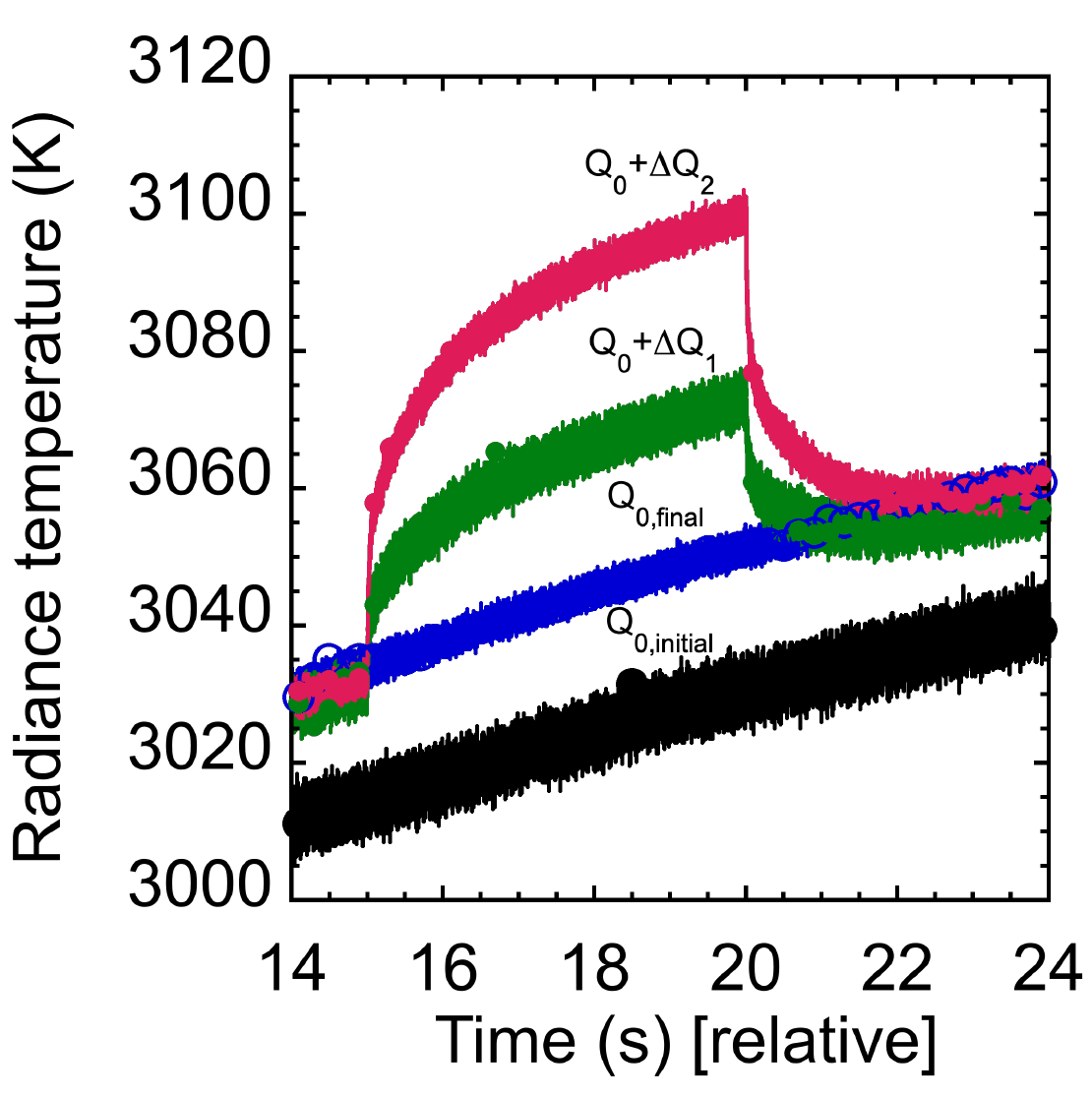}
\caption{Thermograms of the radiance temperature response of W subjected to various power perturbations above the baseline power, $Q_0$, which was used to bring the sample up to a quasi-steady baseline temperature.  The $\Delta$Q perturbations result in a $\Delta T$ above the baseline temperature that is used to calculate the thermal conductivity of the sample.  We note that over the course of collecting a series of data (various thermograms of $\Delta T$ slow temperature rises due to delayed equilibration of the chamber lead to increases in the baseline temperature of the sample ($T_0$).  This is apparent from the difference in radiance temperature of the W from $Q_{0,\mt{initial}}$ to $Q_{0,\mt{final}}$, which were collected using identical powers after 10 successive scans spanning about 30 minutes.  Thus, $\Delta T_{\mt{rad}}$ is determined by interpolating between temperatures at times before and after the change in temperature driven by $\Delta Q$.}
\label{FigData}
\end{center}
\end{figure}

When the sample reaches quasi-steady state at the target temperature by applying some baseline incident power, $Q_0$, we then change the incident laser power by some perturbative $\Delta Q = Q_1-Q_0$ until the sample reaches a new steady state temperature $T_1=\Delta T+T_0$.  Repeating this procedure for multiple $\Delta Q$'s, as shown in Fig.~\ref{FigData}, allows us to determine the slope of the line $\Delta T = A \Delta Q$, which is related to the solution to the steady state heat equation and proportional to $k$, the absorbed power $Q_A$, the laser radius $r_L$, and the boundary conditions.  This problem can be solved analytically assuming a semi-infinite sample, insulative boundary conditions, and a heated region that is circular and on the surface of the sample, given by\cite{carslaw1959ad} $\Delta T = \Delta Q_A/\left(\pi r_L k\right)$ which is the temperature increase in the center of the heated region.  However, given the finite thickness of the sample relative to the laser spot size and thermal gradient inducted by $\Delta Q$ and the radiative losses at the front and back surfaces, we numerically solve the cylindrical heat equation for a finite thickness slab in steady state using an approach detailed previously,\cite{braun2019aa,braun2018aa} modified to account for a top-hat spatial profile of the laser heater. The measured radiance temperatures are corrected to true temperature values via the use of temperature dependent spectral emissivity values at our pyrometer wavelength from literature. Under these quasi-steady-state conditions, we assume the emissivity equals the absorptivity, $A$ to determine the absorbed power at the samples surface at temperature, which we take from the literature for polished W; note, our experiments for thermal conductivity are conducted after several iterations of melting the W surface \textit{in situ} to ensure a specular and clean surface.

In our experiments, the finite size of the sample and radiative boundary conditions (i.e., radiation from surfaces) must be taken into account. For example, in practice, due to spatially dependent temperatures at sample boundaries leading to spatially dependent boundary conductances and additional temperature variations in the chamber and its walls from measurement to measurement (c.f., Fig. 2a), accurate determination of all of the boundary conditions for our precise experiment using exact solution described previously is not possible with any acceptable degree of uncertainty.  Thus, we utilize Stephan Boltzmann's law to calculate the radiative losses emitted from the sample surfaces. The temperature rise on the surface of the sample measured by our single point pyrometer is a result of the absorbed laser power. This incident heat flux is split into two modes of heat transport, conduction and radiation, so that $Q_A = Q_{\mathrm{cond}} + Q_{\mathrm{rad}}$. In order to determine the thermal conductivity of the material we must determine $Q_{\mathrm{cond}}$. Based on the measured true temperature rise, $\Delta T$, we use Stephan Boltzmann's law along with an estimated distribution of radial temperatures to determine the emitted power $Q_{\mathrm{rad}}$. For the small, perturbed temperatures and the relatively fast measurement times of our experiments, we assume the ambient temperature difference between $T_1$ and $T_0$ is negligible. Thus, $Q_{\mathrm{rad}} = \varepsilon \sigma \left(T_1^4-T_0^4\right)$.  To account for these radiative surface losses, we calculate the temperature distributions on the front and back surfaces from solutions to the cylindrical heat equation applied to a finite cylinder with a circular, surface heat source that is detailed in prior works.\cite{olcer1967aa,laughlin1973aa} For solid W, we calculate the radial temperature distribution by assuming a constant temperature (which is the temperature measured by the pyrometer) from the center of the laser spot to $0.55r_L$, and then use that a logarithmic temperature distribution for $\Delta T\left(r\right)$ that decays to 25\% of the center temperature at the outer edge of the 1 cm sample. We assume the same radial temperature distribution for the rear surface temperature, which is opposite the surface on which the laser is applied, only the temperature is $1/3$ that of the front side.  For the molten measurement, in which the sample is only in its molten state in the laser heated volume, the rear temperature is $1/4.5$ that of the front side temperature in the radial portion of the sample.  With these front and rear surface temperature gradients, we then integrate Stephan Boltzmann's law over each surface surface while accounting for the temperature dependent total emissivity\cite{touloukian1970ad} of the material to determine $Q_{\mathrm{rad}}$ emitted by the sample. 

\begin{figure}[t] 
\begin{center}
\includegraphics[scale=.4]{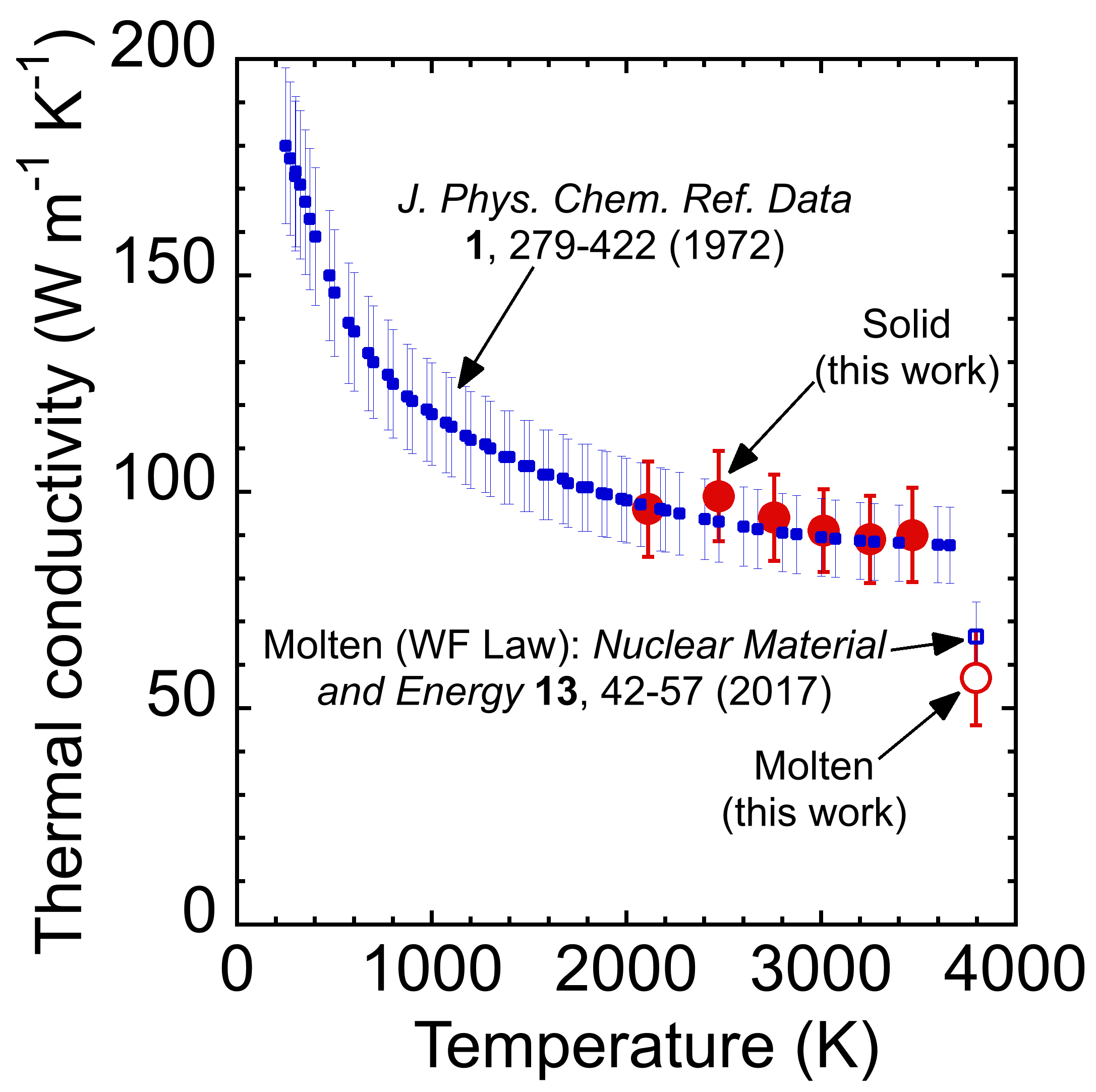}
\caption{Measured thermal conductivity of solid and molten W from this work (fill and open circles, respectively) compared to accepted values from the literature for both solid\cite{ho1972aa} and molten states.\textit{et al.}\cite{tolias2017aa} The previously reported accepted value for the thermal conductivity of molten W from Tolias \textit{et al.}\cite{tolias2017aa} is derived from the Wiedemann-Franz law applied to electrical resistivity data.}
\label{FigTCWtemp}
\end{center}
\end{figure}

The measured thermal conductivity of the W sample as a function of temperature is shown in Fig.~\ref{FigTCWtemp} up to and above its melting temperature.  For the solid phase, we plot the recommended values for the thermal conductivity of W tabulated in the literature,\cite{ho1972aa} which agree well with our measurements.  We are not aware of any prior measurements of the thermal conductivity of molten W that were not derived from electrical resistivity measurements via the Wiedemann-Franz Law, and thus our measurement offers a unique opportunity to validate the Wiedemann-Franz Law for molten tungsten.  Tolias \textit{et al.}\cite{tolias2017aa,pottlacher2010aa} and Pottlacher report recommended values for the electrical resistivity of molten W at temperatures near the phase transition of $\rho = 134$ $\mu\Omega$-$\mt{cm}$ and  $\rho = 137$ $\mu\Omega$-$\mt{cm}$, respectively.  Using $L_0 = 2.443\times10^{-8}$ W $\Omega$ K$^{-2}$ yields a thermal conductivity of $68-70$ W m$^{-1}$ K$^{-1}$.  Tolias \textit{et al.}\cite{tolias2017aa} further expounds upon their molten W thermal conductivity report to recommend a value of $66.6\pm8.2$ W m$^{-1}$ K$^{-1}$ and discontinuity in $k$ across the solid-liquid phase transition of $20.4$ W m$^{-1}$ K$^{-1}$.  This is in agreement with our measured value of $57.0\pm11.0$ W m$^{-1}$ K$^{-1}$ with a similar reduction in thermal conductivity from the solid to molten state within our experimental uncertainty.  Taken together, our measurements strongly support the validity of the Wiedemann-Franz Law and the use of $L_0$ applied to W in its molten state.

\begin{figure}[t] 
\begin{center}
\includegraphics[scale=.4]{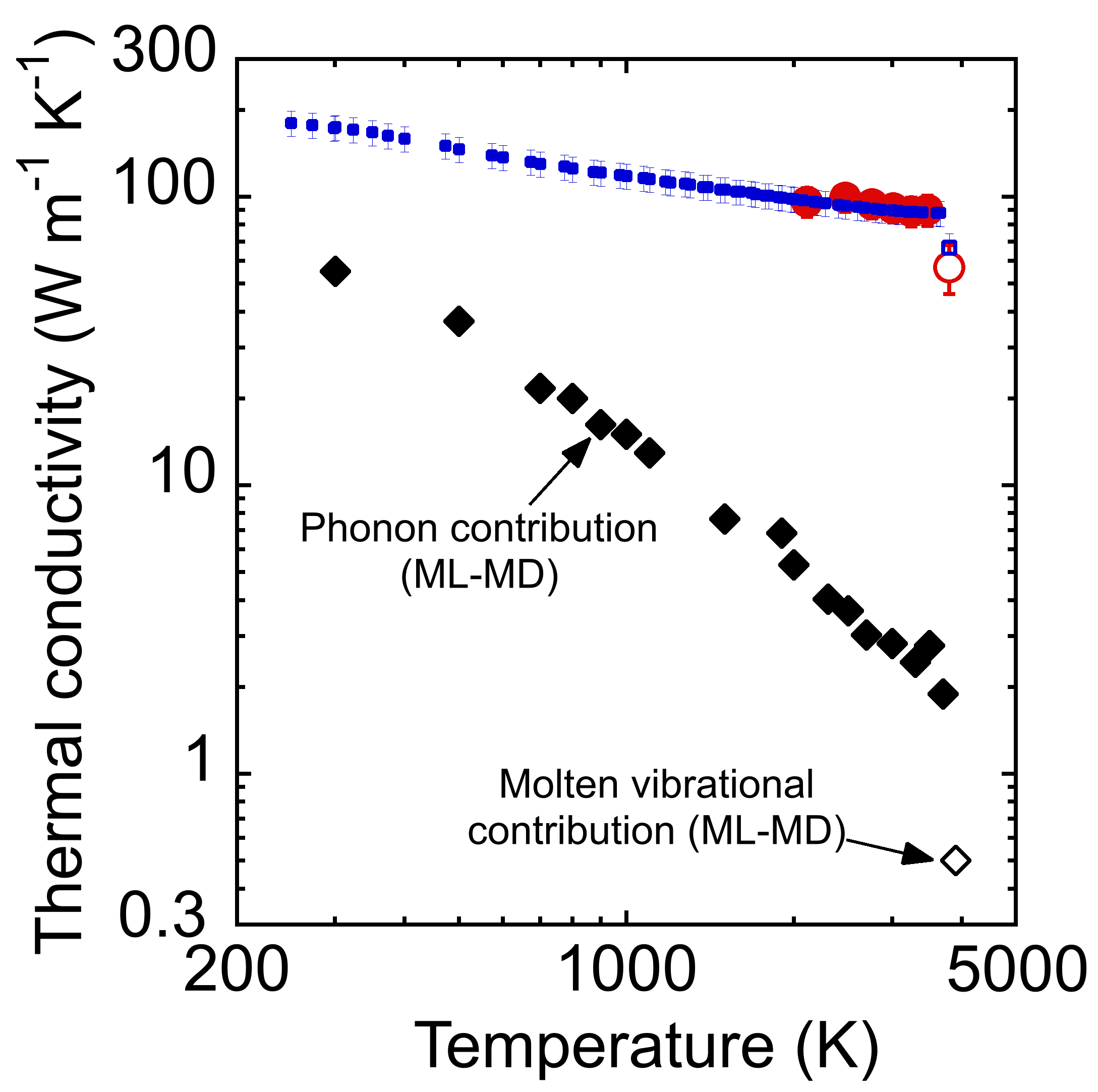}
\caption{ML-MD simulated vibrational thermal conductivities of solid (phonons) and molten (vibrations) W compared to data presented in Fig.~\ref{FigTCWtemp} (data symbols the same as in Fig.~\ref{FigTCWtemp}). The vibrational component to the thermal conductivity of W only contributes $<5\%$ to the total thermal conductivity above 2000 K with the phononic contribution decreasing with increasing temperature up to the melting temperature. In the molten state, the vibrational thermal conductivity is $\sim0.5$ W m$^{-1}$ K$^{-1}$, contributing $<1\%$ to the thermal conductivity. At these elevated temperatures above $T_{\mt{melt}}/2$, the vibrational contribution is effectively suppressed leading to the thermal conductivity of W being dominated by electrons, thus resulting in the Wiedemann-Franz Law proving an accurate route to calculate the thermal conductivity.}
\label{FigTCWTempwMDlog}
\end{center}
\end{figure}

We further validate our assertion that the Wiedemann-Franz Law is suitable to calculate the thermal conductivity of high temperature W by determining the lattice contribution to the total thermal conductivity of W through \textit{ab initio} molecular dynamics (MD) simulations, where we develop a machine-learned (ML) potential for W that is uniformly accurate over the entire range of temperatures from room temperature up through the melting point.  These ML-MD simulated phonon thermal conductivities of W are shown in Fig.~\ref{FigTCWTempwMDlog}, and only contribute $<5\%$ to the total thermal conductivity of W at these high temperatures with the phononic contribution decreasing with increasing temperature up to the melting temperature. In the molten state, the vibrational thermal conductivity is $\sim0.5$ W m$^{-1}$ K$^{-1}$, contributing $<1\%$ to the thermal conductivity. While this relatively negligible phononic/vibrational contribution is not the case at lower temperatures (i.e., room temperature) where the phonon contribution to the thermal conductivity of W can contribute $\sim30\%$, as predicted by our ML-MD simulations and consistent with prior works,\cite{chen2019aa} at these elevated temperatures above $T_{\mt{melt}}/2$, the vibrational contribution is effectively suppressed leading to the thermal conductivity of W being dominated by electrons, thus resulting in the Wiedemann-Franz Law proving an accurate route to calculate the thermal conductivity.

In conclusion, we measure the thermal conductivity of solid and molten tungsten using Steady State Temperature Differential Radiometry (SSDTR).  We demonstrate that the thermal conductivity can be well described by application of Wiedemann-Franz Law using the Lorentz constant $L_0$ to electrical resistivity data.  We further support this conclusion using \textit{ab initio} molecular dynamics simulations with a machine-learned potential that is uniformly accurate over the entire range of temperatures from room temperature up through the melting point.  Our results show that at these high temperatures, the vibrational contribution to thermal conductivity is negligible compared to the electronic component, thus validating the validity of Wiedemann-Franz Law to capture the electronic thermal conductivity of W at $T_{\mt{melt}}/2$, including in its molten state.

Our conclusions are enabled through the development of SSTDR as a measurement technique for determination of thermal conductivity, which allows for a measure of thermal properties without relying on interpretation of electrical resistances.  This technique could offer an approach to measure the thermal properties of electrical insulating materials at high temperatures, including during melting and in their molten states.  With integration of additional spatial and spectral pyrometric sensors with increased temporal resolution into this apparatus, this technique can be extended to measure heat capacity, emissivity, and latent heat of materials at high temperatures and during solid-liquid transitions.

\section*{Acknowledgements}
We appreciate support from the Office of Naval Research, Grant No.~N00014-22-1-2139.  P.E.H.~acknowledges support from the Alexander von Humboldt Foundation for supporting his work at the European Commission, Joint Research Centre in Karlsruhe, Germany.

\end{document}